\documentclass[prd,reprint,showpacs,showkeys]{revtex4-1}
\usepackage{amsfonts}
\usepackage{amssymb}
\usepackage{amsmath}
\usepackage{graphicx}
\usepackage[font={footnotesize,it}]{caption}

\begin{document}

\title{Morse simulation of the Global Monopole equation in flat spacetime}
\author{S. Habib Mazharimousavi}
\email{habib.mazhari@emu.edu.tr}
\author{M. Halilsoy}
\email{mustafa.halilsoy@emu.edu.tr}
\affiliation{Department of Physics, Eastern Mediterranean University, Gazima\~{g}usa,
Turkey. }
\date{\today }

\begin{abstract}
We show that a Morse type potential simulates an analytic solution for the
highly non-linear global monopole field equation in three and higher
dimensional flat spacetimes. Owing to the fact that in the flat space limit
the similar equation remains intact we wish to borrow the curved space
terminology of global monopole also in flat spacetime. This may provide a
compelling example that can be used effectively in different non-linear
theories such as flat space $\phi ^{4},$ as well as in curved spacetimes.
\end{abstract}

\maketitle

\section{Introduction}

In the absence of exact solutions non-linear differential equations of field
theory manifest features that they can be simulated by some known
potentials. Linear equations with complicated potentials also can be treated
similarly. In this regard we recall the $H-$atom endowed with non-linear
electromagnetic field due to Born and Infeld \cite{BI}. In this approach
basic idea was to eliminate the divergences due to point charges. The
resulting regular electromagnetic potential expected to contribute to the
solution of Schr\"{o}dinger equation and the ground state energy level was
obtained remarkably as a simulation \cite{MH} of the Morse potential \cite%
{Morse}. Being motivated by that study we wish to employ the same method to
another well-known differential equation representing an analog of global
monopole in flat spacetime. Global monopoles are topological structures that
emerge from spontaneous symmetry breaking and are believed to exist from the
big bang. Such a monopole was considered first by Barriola and Vilenkin \cite%
{BV} as source of gravity which may go to the extend of forming black holes.
In $3+1-$dimensions the symmetry that has been broken is $O(3)$ which
ultimately reduces to $U(1)$. In this approach \cite{BV} the scalar field
triplet is given by $\phi ^{a}=\eta f\left( r\right) \frac{x^{a}}{r},$ where 
$a=1,2,3$ are the gauge parameters, $\eta $ is the monopole charge and $%
f\left( r\right) $ is the function to be determined. Unfortunately the
differential equation satisfied by $f\left( r\right) $ doesn't admit an
exact analytical solution. The asymptotic behaviors satisfied by $f\left(
r\right) $ are such that $f\left( r\right) \rightarrow 0$ for $r\rightarrow
0,$ and $f\left( r\right) \rightarrow 1$ with $f^{\prime }\left( r\right) =0$
for $r\rightarrow \infty $. Numerical integration of the resulting
differential equation is at our disposal and our aim is to simulate such a
global monopole equation by using a Morse type potential. Before extending
our formalism to higher dimensions we consider the $2+1-$dimensional
analogue of the $3+1-$dimensional global monopole. In $2+1-$dimensions the
gauge group to be broken becomes now $O(2)$ so that the scalar doublet is
represented by $\phi ^{a}=\eta f\left( r\right) \frac{x^{a}}{r},$ where $%
a=1,2.$ The resulting differential equation for $f\left( r\right) $ in $2+1-$%
dimensions is as difficult as in the $3+1-$dimensional case. As a matter of
fact asymptotically the resulting equation is reminiscent of a minor
variation of the Third Painlev\'{e} transcendent \cite{PL} encountered in
the study of cylindrical gravitational waves \cite{Chandra}. This
oscillatory solution, however, doesn't represent the monopole character but
yet arises as a distinct solution to the same differential equation
satisfied by the monopole. In other words both types of solutions arise from
the same differential equation with different initial conditions. Morse
potential fitting to the global monopole equation in $2+1-$dimensions is
observed to work perfectly. Going to the lower dimension $1+1$, the
differential equation satisfied by $f\left( r\right) $ is solved exactly as $%
f\left( r\right) =\tanh \left( \frac{r}{\sqrt{2}}\right) .$ This is,
however, a singlet field and it can't be interpreted as a global monopole
since there is no symmetry breaking in the usual sense. Reflection symmetry
is the only available symmetry to be in $1-$dimensional space. Next, we
interpolate our analysis to higher dimensions $D\geq 4$ and repeat our
method of Morse simulation. We observe that as $D$ gets higher the fitting
of our Morse type potential becomes less reliable in comparison with the
dimensions $2+1$ and $3+1$.

\section{Global monopole equation in $2+1-$dimensions}

Let's start with the Lagrangian of a global-monopole in $2+1-$dimensional
flat, spherically symmetric spacetime which is given by%
\begin{equation}
\mathcal{L}^{field}=-\frac{1}{2}\partial _{\mu }\phi ^{a}\partial ^{\mu
}\phi ^{a}-\frac{1}{4}\lambda \left( \phi ^{a}\phi ^{a}-\eta ^{2}\right)
^{2}.
\end{equation}%
Here $a=\left( 1,2\right) $, $\lambda $ is a coupling constant, $\eta $ is
the monopole parameter and%
\begin{equation}
\phi ^{a}=\eta f\left( r\right) \frac{x^{a}}{r},
\end{equation}%
for $x^{1}=r\cos \theta $ and $x^{2}=r\sin \theta .$ To find the field
equation for $f\left( r\right) $ we express the field Lagrangian in terms of 
$f\left( r\right) $ only, i.e., 
\begin{equation}
\mathcal{L}^{field}=-\frac{\eta ^{2}}{2}\left( f^{\prime 2}+\frac{f^{2}}{%
r^{2}}\right) -\frac{\lambda \eta ^{4}}{4}\left( f^{2}-1\right) ^{2}.
\end{equation}%
Now, variation of the action with respect to $f$ yields%
\begin{equation}
\left( rf^{\prime }\right) ^{\prime }-\frac{f}{r}-\lambda \eta ^{2}rf\left(
f^{2}-1\right) =0
\end{equation}%
or in more convenient form%
\begin{equation}
f^{\prime \prime }+\frac{1}{r}f^{\prime }+\left( \frac{1}{\delta ^{2}}-\frac{%
1}{r^{2}}\right) f=\frac{1}{\delta ^{2}}f^{3}
\end{equation}%
in which a prime stands for the derivative with respect to $r$ and $\delta =%
\frac{1}{\eta \sqrt{\lambda }}$ is the size of the global monopole in flat
spacetime. We note that a scaling of the form $r\rightarrow r\delta ,$
absorbs the parameter, hence without loss of generality one can set $\delta
=1.$ Note also that the term $-\frac{f}{r^{2}}$ is as a result of symmetry
breaking.

Let us note that Eq. (5) admits an asymptotic solution in analogy with the
cylindrical gravitational waves expressed exactly in terms of Painlev\'{e}
type III \cite{Chandra}. Our case turns out to be%
\begin{multline}
f\left( r\right) \sim \frac{1}{\sqrt{r}}\cos \left( r+\frac{3}{8}\ln
r+\alpha _{0}\right) - \\
\frac{1}{32r^{3/2}}\cos 3\left( r-\frac{3}{8}\ln r+\alpha _{0}\right)
+O\left( r^{-5/2}\right)
\end{multline}%
(for $\alpha _{0}=$constant) which is comparable with the asymptotic
solution given in \cite{Chandra}.

In Fig. 1 we plot the solution of latter equation for $\delta =1$ and $%
f\left( 0\right) =0.$ The solution is oscillatory for $f^{\prime }\left(
0\right) <0.583189028596$, asymptotes to one for $f^{\prime }\left( 0\right)
=0.583189028596$ and diverges for $f^{\prime }\left( 0\right)
>0.583189028596.$ Interestingly the decisive factor between the two types of
solutions is the initial slope of the curve to start with. Below certain
value we have oscillations while above that we have the frozen global
monopole solution.

\begin{figure}[tbp]
\includegraphics[width=70mm,scale=0.7]{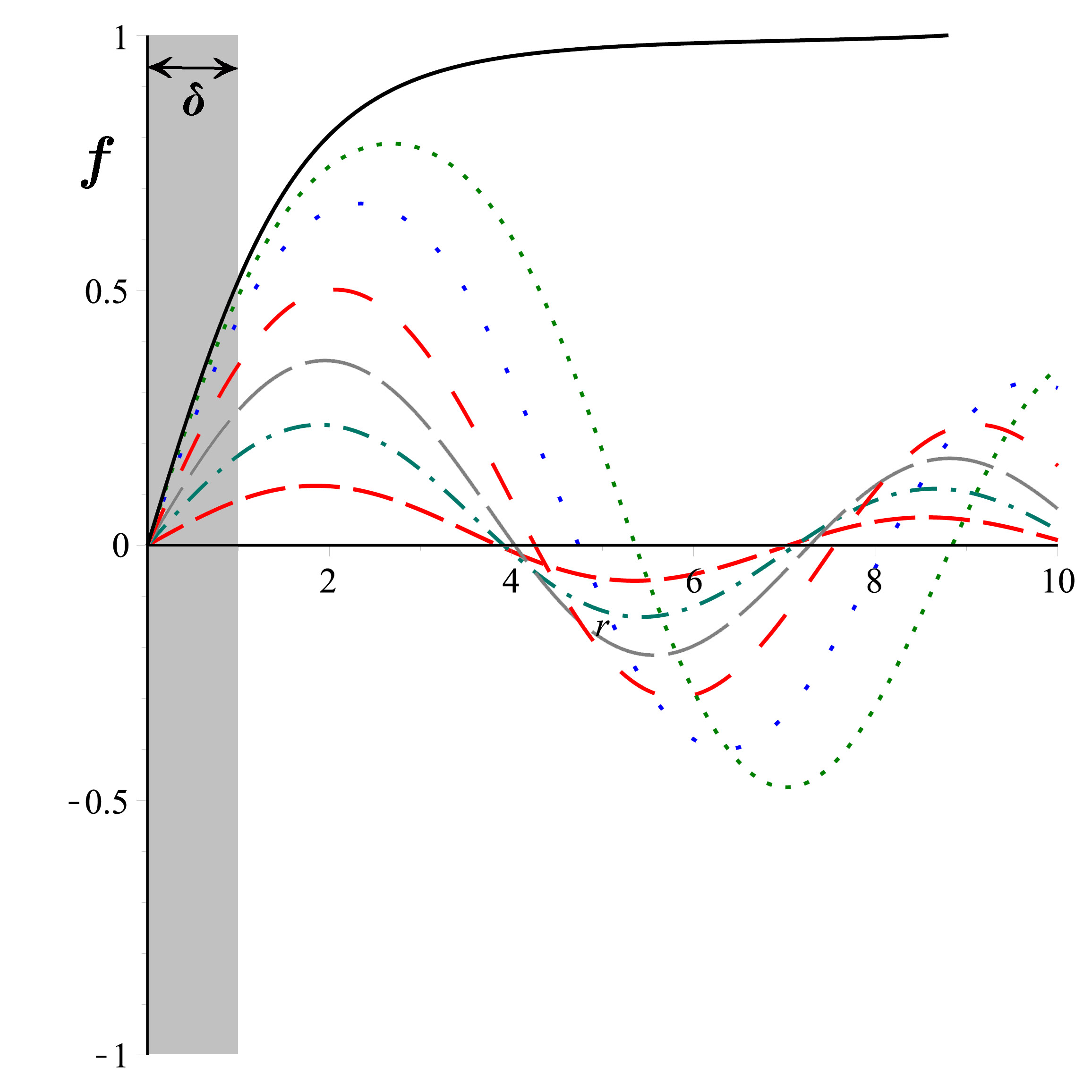} %
\captionsetup{justification=raggedright, singlelinecheck=false}
\caption{A plot of the solution to the field equation (5) $f\left( r\right) $
with respect to $r$ for $\protect\delta =1$ in flat spacetime. The initial
conditions are set as $f\left( 0\right) =0$ and $f^{\prime }\left( 0\right)
=0.583189029,$ $0.55,0.5,0.4,0.3,0.2$ and $0.1$ from the top/Black-Solid to
the bottom/Red-long Dash. Note that the shaded region corresponds to the
inner region of global-monopole. We add also that depending on the initial
slope Eq. (5) admits a wave like solution which is apart from the monopole.
For completeness we have shown the latter as well.}
\end{figure}
\begin{figure}[tbp]
\includegraphics[width=70mm,scale=0.7]{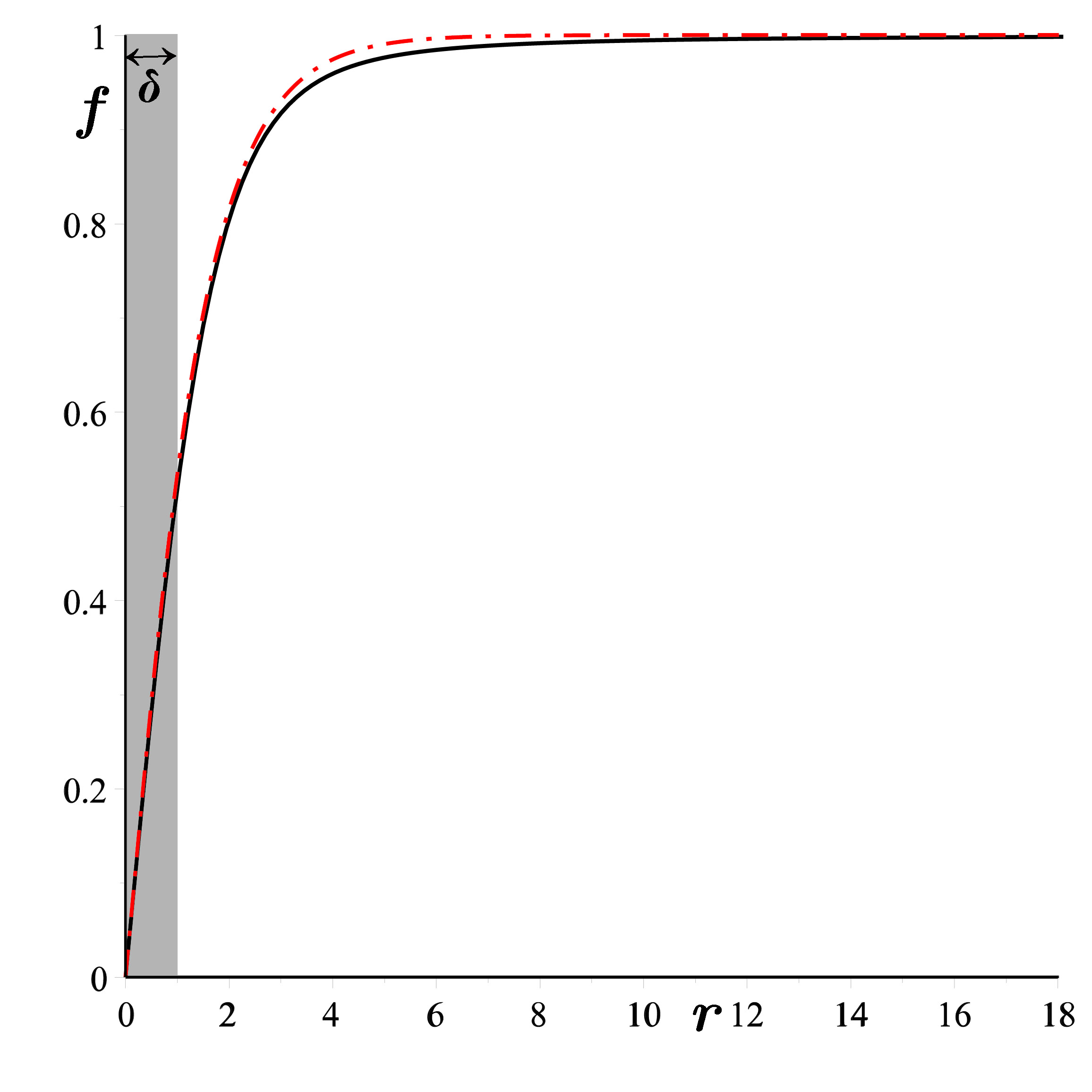} %
\captionsetup{justification=raggedright, singlelinecheck=false}
\caption{A plot (black-solid) of the solution to the field equation (5) $%
f\left( r\right) $ with respect to $r$ for $\protect\delta =1$ in flat
spacetime for the initial conditions are set as $f\left( 0\right) =0$ and $%
f^{\prime }\left( 0\right) =0.583189029,$ and the Morse function (red-dash).}
\end{figure}
In Fig. 2 we plot the numerical solution of Eq. (5) which represents a
global monopole with $\delta =1$ and the Morse type function%
\begin{equation}
f\left( r\right) \simeq 1-\left( f_{0}^{\prime }-1\right) e^{-2r}+\left(
f_{0}^{\prime }-2\right) e^{-r}
\end{equation}%
in which $f_{0}^{\prime }=0.583189028596$. Obviously $f\left( r\right) $
satisfies the conditions at $r=0$ and $r\rightarrow \infty .$

\section{Global monopole equation in all dimensions}

In higher dimensional spherical symmetric flat spacetime the action of a
global monopole field is given by (1) and (2), with $a=\left(
1,2,...,D-1\right) $ and 
\begin{eqnarray}
x^{1} &=&r\cos \theta _{1}  \notag \\
x^{2} &=&r\sin \theta _{1}\cos \theta _{2}  \notag \\
x^{3} &=&r\sin \theta _{1}\sin \theta _{2}\cos \theta _{3}  \notag \\
&&.  \notag \\
&&. \\
&&.  \notag \\
x^{D-2} &=&r\sin \theta _{1}...\sin \theta _{D-3}\cos \theta _{D-2}  \notag
\\
x^{D-1} &=&r\sin \theta _{1}...\sin \theta _{D-3}\sin \theta _{D-2}  \notag
\end{eqnarray}%
where $\theta _{D-2}\in \left[ 0,2\pi \right] $ and $\theta _{k}\in \left[
0,\pi \right] $ with $k=1,2,...,D-3.$ The arbitrary dimensional action can
be expressed in the form%
\begin{equation}
\mathcal{L}^{field}=-\frac{\eta ^{2}}{2}\left( f^{\prime 2}+\frac{D_{2}f^{2}%
}{r^{2}}\right) -\frac{\lambda \eta ^{4}}{4}\left( f^{2}-1\right) ^{2}
\end{equation}%
in which $D_{2}=D-2$ with $D$ the dimension of the spacetime. The field
equation following the above Lagrangian is found to be%
\begin{equation}
f^{\prime \prime }+\frac{D_{2}}{r}f^{\prime }+\left( \frac{1}{\delta ^{2}}-%
\frac{D_{2}}{r^{2}}\right) f=\frac{1}{\delta ^{2}}f^{3}
\end{equation}%
with $\delta =\frac{1}{\eta \sqrt{\lambda }}$ the size of the higher
dimensional global monopole. In Fig. 3 we plot the global monopole solution
of the latter equation for $D=2$ (top/black-long dash), $D=3,4,5,6$ and $%
D=10 $ (bottom/purple-dot). The case when the dimension is two i.e., $D=2$
is analytically solvable. The field equation for $D=2$ and $\delta =1$ can
be written as%
\begin{equation}
f^{\prime \prime }+f\left( 1-f^{2}\right) =0
\end{equation}%
with a first integral%
\begin{equation}
f^{\prime 2}=\frac{1}{2}f^{4}-f^{2}+C
\end{equation}%
in which $C$ is an integration constant. 
\begin{figure}[tbp]
\includegraphics[width=70mm,scale=0.7]{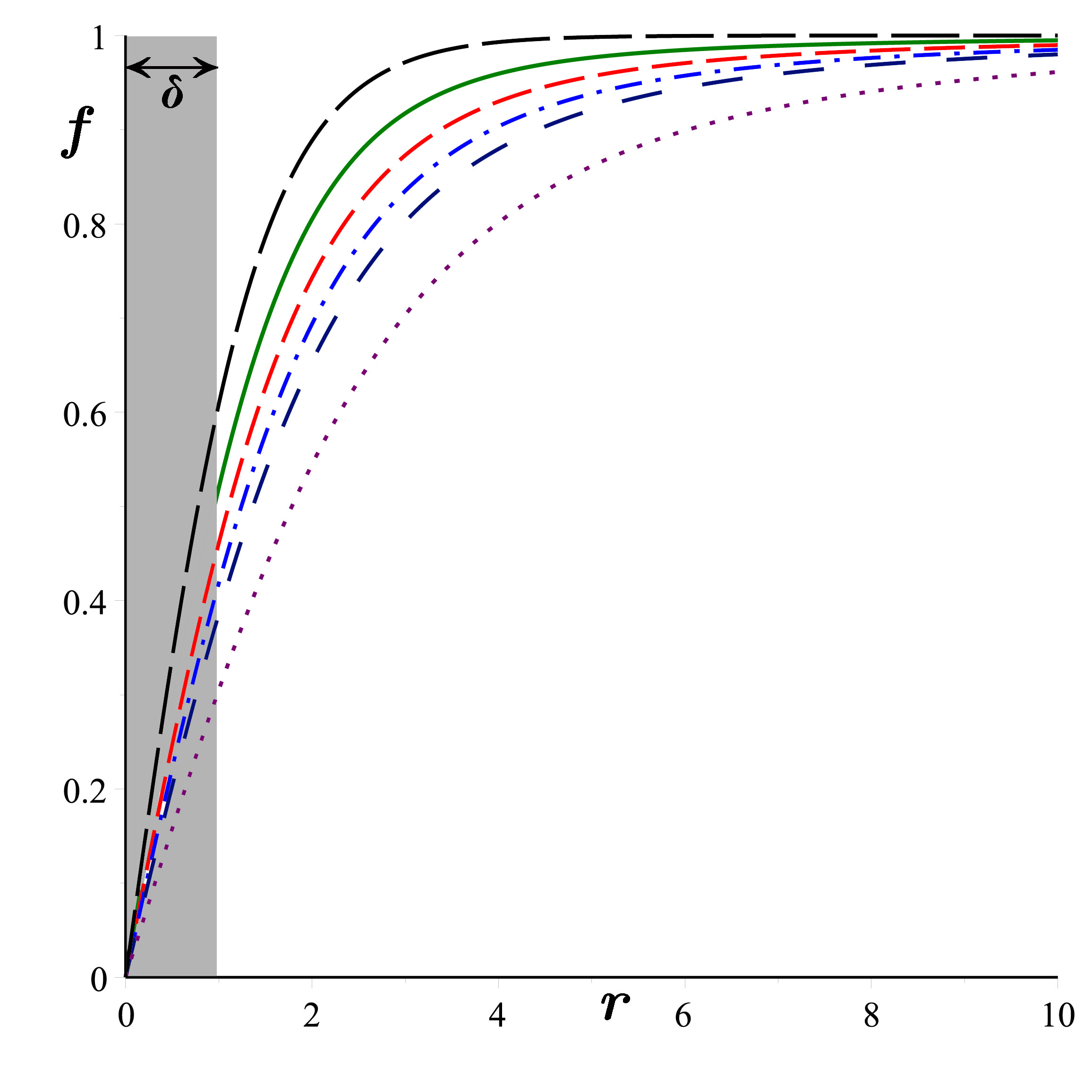} %
\captionsetup{justification=raggedright, singlelinecheck=false}
\caption{Plot of numerical global monopole solution to the general field
equation (10) for dimensions $D=2,3,4,5,6,10$ from the top to the bottom. }
\end{figure}
\begin{figure}[tbp]
\includegraphics[width=70mm,scale=0.7]{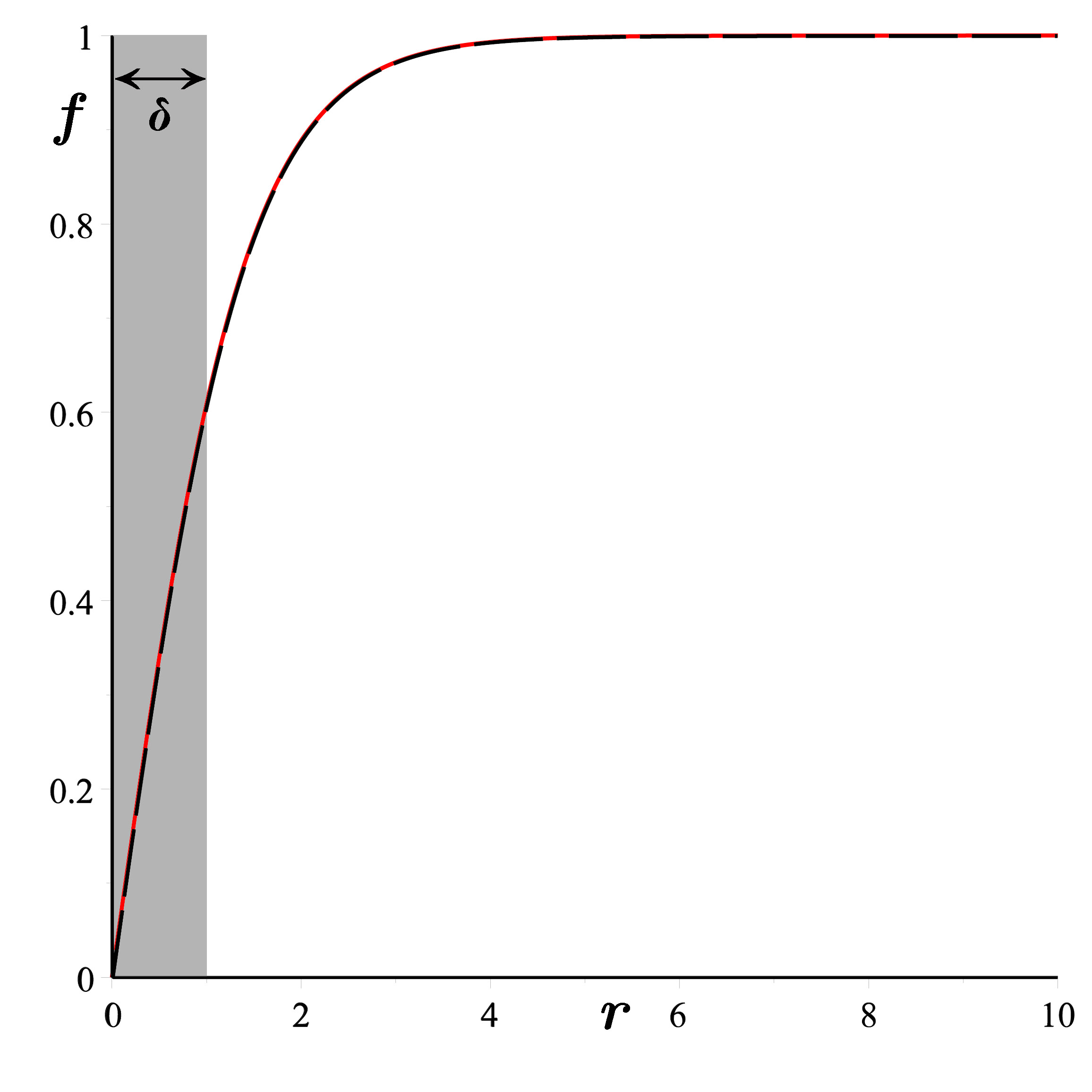} %
\captionsetup{justification=raggedright, singlelinecheck=false}
\caption{Plot of exact solution for the global monopole function $f$ and the
numeric solution for $D=2.$ The two solutions agree highly which can not be
distinguished from each other.}
\end{figure}
As we are looking for a global monopole type solution, we impose the
constraint that when $r\rightarrow \infty $, $f\rightarrow 1$ and $f^{\prime
}\rightarrow 0.$ This in turn implies that $C=\frac{1}{2}.$ Upon considering 
$C$ in (12) we obtain%
\begin{equation}
f^{\prime 2}=\frac{1}{2}\left( f^{2}-1\right) ^{2}.
\end{equation}%
Furthermore the global monopole field admits a positive derivative and is
less than one which imply%
\begin{equation}
f^{\prime }=\frac{1}{\sqrt{2}}\left( 1-f^{2}\right) .
\end{equation}%
This final equation admits an exact solution given by%
\begin{equation}
f=\tanh \left( \frac{r}{\sqrt{2}}+\tilde{C}\right)
\end{equation}%
in which $\tilde{C}$ is an integration constant. Imposing $f\left( 0\right)
=0,$ one finds $\tilde{C}=0$ which makes the final solution simply as%
\begin{equation}
f=\tanh \left( \frac{r}{\sqrt{2}}\right) .
\end{equation}%
In Fig. 4 we plot the exact solution (16) and the numerical solution found
from the field equation. We see that they agree to high precision. This
accuracy guides our numerical method in $D=3$ also to the higher dimensions.

Similar to $D=3,$ for $D\geq 4$ we can simulate the solution with a Morse
type function given by%
\begin{equation}
f\left( r\right) \simeq 1-\left( \frac{f_{0}^{\prime }}{\kappa }-1\right)
e^{-2\kappa r}+\left( \frac{f_{0}^{\prime }}{\kappa }-2\right) e^{-\kappa r}
\end{equation}%
in which $f_{0}^{\prime }$ is the slope of the solution at $r=0$ and $\kappa 
$ is an adjusting parameter. In Table. 1 we give $f_{0}^{\prime }$ for
dimensions $D\geq 4$ and in Fig. 5 we plot the numerical solution with the
simulated function for $D=4.$

\begin{table}[th]
\caption{$f_{0}^{\prime }$ for different dimensions.}
\centering
\begin{tabular}{cc}
\hline\hline
$D$ & $f_{0}^{\prime }$ \\[1ex] \hline
$2$ & $0.707106680906$ \\ 
$3$ & $0.583189028596$ \\ 
$4$ & $0.506042540468$ \\ 
$5$ & $0.4525666770699$ \\ 
$6$ & $0.4128548316012$ \\ 
$10$ & $0.3188665791567$ \\ \hline
\end{tabular}
\end{table}
\begin{figure}[tbp]
\includegraphics[width=70mm,scale=0.7]{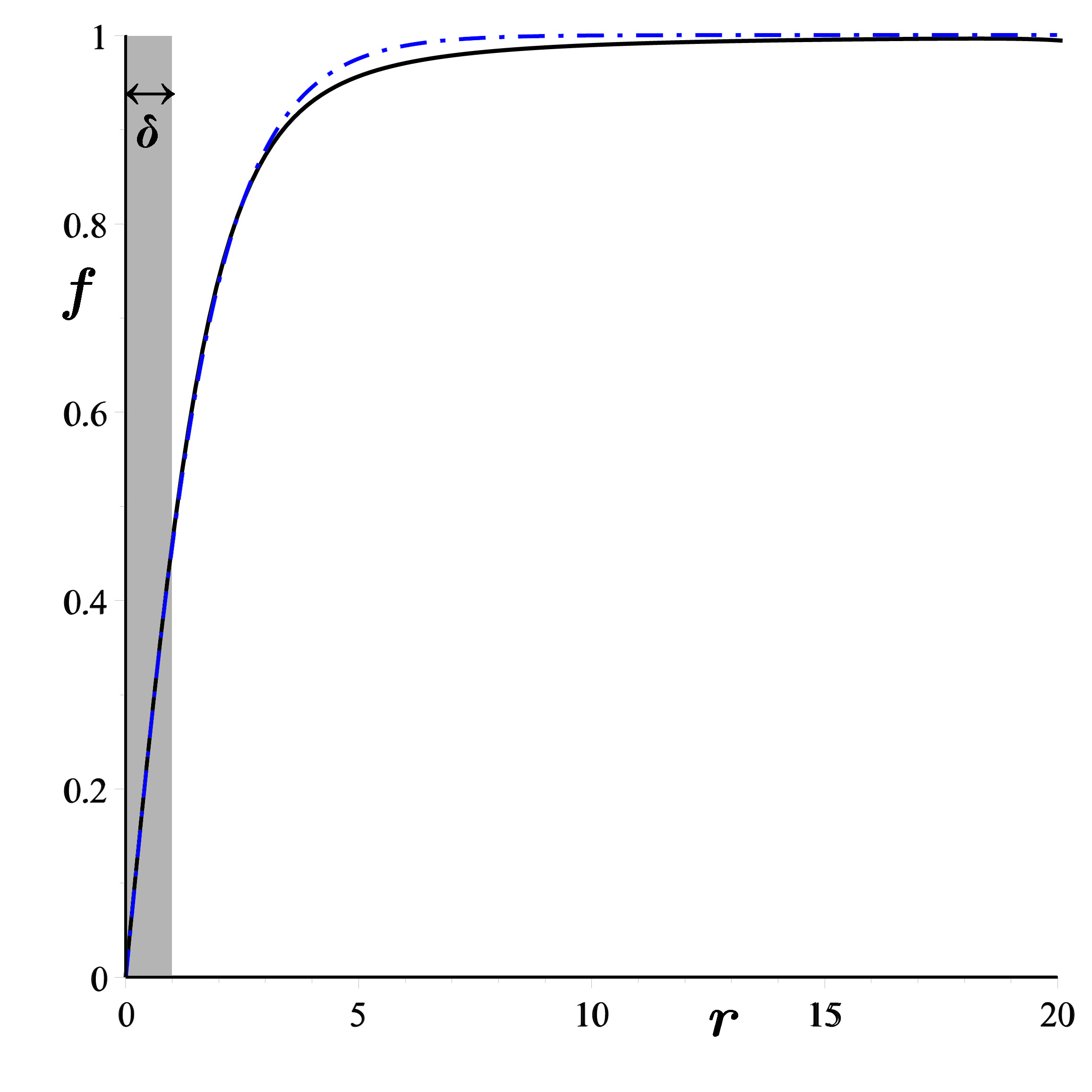} %
\captionsetup{justification=raggedright, singlelinecheck=false}
\caption{Plot of simulating Morse function (blue-dash) for the global
monopole function $f$ and the numeric solution (black-solid) for $D=4.$ In
the Morse function we set $\protect\kappa =0.8.$ The two solutions agree in
both small and large $r$ to high accuracy.}
\end{figure}

\section{Why the Morse type function?}

Having the solution of Eq. (10) simulated by the Morse type function is not
a coincidence. The initial form of the Morse potential can be written as 
\cite{Morse}%
\begin{equation}
V_{Morse}\left( r\right) =D_{e}\left( 1-e^{-a\left( r-r_{e}\right) }\right)
^{2}
\end{equation}%
in which $D_{e}$, $a$ and $r_{e}$ are free positive parameters of the
potential. This function has a minimum at $r=r_{e}$ and for large $r$ it
approaches to $D_{e}.$ Setting $D_{e}=1$ and $r_{e}=0$ we find a function
whose asymptotic behaviours are as we are expecting for $f\left( r\right) $
in Eq. (10) i.e., $\lim_{r\rightarrow 0}V_{Morse}\left( r\right) =0$ and $%
\lim_{r\rightarrow \infty }V_{Morse}\left( r\right) =1.$\ Furthermore, for
large $r$ i.e., $\frac{r}{\delta }\gg 1,$ Eq. (10) becomes independent of
dimensions and therefore the solution must be asymptotically the same as the
exact solution of (10) in $D=2$ given by (16). Hence we get%
\begin{equation}
\lim_{r\rightarrow \infty }f=\lim_{r\rightarrow \infty }\tanh \left( \frac{r%
}{\zeta }\right) 
\end{equation}%
where $\zeta $ is a positive constant which could be set according to the
dimensions. The asymptotic behavior of the right hand side in (19)
surprisingly is of the Morse type function i.e., for large $r$%
\begin{equation}
\tanh \left( \frac{r}{\zeta }\right) \sim 1-2e^{\frac{-2r}{\zeta }}+2e^{%
\frac{-4r}{\zeta }}.
\end{equation}%
Finally, for small $r$ i.e., $r/\delta \ll 1$ the field equation (10) becomes%
\begin{equation}
r^{2}f^{\prime \prime }+D_{2}rf^{\prime }-D_{2}f=0
\end{equation}%
with a solution given by%
\begin{equation}
f\left( r\right) =C_{1}r+C_{2}r^{-D_{2}}
\end{equation}%
in which $C_{1}$ and $C_{2}$ are constants. A regular solution at $r=0$
requires $C_{2}=0$ and the asymptotic solution for small $r$ becomes $%
f\left( r\right) =C_{1}r.$ From the other hand the asymptotic behaviour of
the Morse potential (18) with $D_{e}=1$ and $r_{e}=0$ near $r=0$ becomes%
\begin{equation}
\lim_{r\rightarrow 0}V_{Morse}\left( r\right) \sim ar
\end{equation}%
which is same as $f\left( r\right) .$ In conclusion the Morse type function
in both limits i.e., $r\rightarrow 0$ and $r\rightarrow \infty $ matches
very well with the solution of the field equation (10). As we expect from
the Fig. 5, the maximum agreement occurs in these limits.

\section{Conclusion}

Irrespective of dimensions ($D\geq 3$) the geometry of a global monopole
emerges complicated enough to be tackled with, even in a flat spacetime. The
reason is the transcendental differential equation satisfied by the global
monopole function $f\left( r\right) .$ Asymptotically the equation satisfied
by $f\left( r\right) $ has similarity with the Painlev\'{e} type III, whose
behavior can be studied numerically. We have shown that the numerical plot
of the function can be simulated by a Morse type function to great accuracy.
It is shown that the accuracy of our simulation works better in lower
dimensions such as $D=3,4$. $D=2$ is the only available case that admits
exact integration, however, its significance as a monopole remains
questionable. Recalling that a global monopole constitutes a particular case
of $\phi ^{4}$ theories similar simulations may find application also in the
latter theories. Further, question can naturally be raised about the
importance of such a simulation advocated herein: the physics of particles
in the field of a monopole becomes tractable through the Morse function. Our
project next is to extend the idea to global monopole in a curved spacetime.

\bigskip

\bigskip


\begin{thebibliography}{9}
\bibitem{BI} M. Born and L. Infeld, Proc. Roy. Soc, A \textbf{144}, 425
(1934).

\bibitem{MH} S. H. Mazharimousavi and M. Halilsoy, Found Phys. \textbf{42},
524 (2012).

\bibitem{Morse} P. M. Morse, Phys. Rev. \textbf{34}, 57 (1929).

\bibitem{BV} M. Barriola and A. Vilenkin, Phys. Rev. Lett. \textbf{63}, 341
(1989).

\bibitem{PL} E. L Ince, "Ordinary Differential Equations", Published by
Dover, New York, (1956).

\bibitem{Chandra} S. Chandrasekhar, Proc. Roy. Soc. A \textbf{408}, 209
(1986).
\end{thebibliography}
\end{document}